\def\year{2020}\relax
\documentclass[letterpaper]{article} % DO NOT CHANGE THIS
\usepackage{aaai20}  % DO NOT CHANGE THIS
\usepackage{times}  % DO NOT CHANGE THIS
\usepackage{helvet} % DO NOT CHANGE THIS
\usepackage{courier}  % DO NOT CHANGE THIS
\usepackage[hyphens]{url}  % DO NOT CHANGE THIS
\usepackage{graphicx} % DO NOT CHANGE THIS
\usepackage{graphicx}  % DO NOT CHANGE THIS
\usepackage{amssymb}
\usepackage{amsmath}
\usepackage{rotating}
\usepackage{dsfont}

\newcommand{\Prv}{\mathit{Prv_{PA}}}
\newcommand{\Opp}{\mathit{Opposite}}
\newcommand{\QB}[1]{\ulcorner #1 \urcorner}
\newcommand{\Con}{\mathit{Con_{PA}}}

\newcommand{\Paragraph}[1]{\noindent\textbf{#1}}

\title{Towards Concise, Machine-discovered Proofs of G\"{o}del's Two Incompleteness Theorems}
\author{Elijah Malaby, Bradley Dragun, John Licato\\
Advancing Machine and Human Reasoning (AMHR) Lab\\
Department of Computer Science and Engineering\\
University of South Florida
}
\begin{document}

\maketitle

\begin{abstract}
There is an increasing interest in applying recent advances in AI to automated reasoning, as it may provide useful heuristics in reasoning over formalisms in first-order, second-order, or even meta-logics. To facilitate this research, we present MATR, a new framework for automated theorem proving explicitly designed to easily adapt to unusual logics or integrate new reasoning processes. MATR is formalism-agnostic, highly modular, and programmer-friendly. We explain the high-level design of MATR as well as some details of its implementation. To demonstrate MATR's utility, we then describe a formalized metalogic suitable for proofs of G\"odel's Incompleteness Theorems, and report on our progress using our metalogic in MATR to semi-autonomously generate proofs of both the First and Second Incompleteness Theorems.
\end{abstract}

%%%%%%%%%%%%%%%%%%%%%%%%%%%%%%%%%%%%%%%%%%%%%%%%%%%%%%%%%%%%
%%%%%%%%%%%% PAGE LIMIT: 6, INCLUDING CITATIONS %%%%%%%%%%%%
%%%%%%%%%%%%%%%%%%%%%%%%%%%%%%%%%%%%%%%%%%%%%%%%%%%%%%%%%%%%

\section{Introduction}
\label{sec:intro}

\noindent An emerging body of literature seeks to apply the recent advances of machine learning and deep networks to the field of automated theorem proving. For example, given a partially completed deductive proof, deciding which inference rules to apply might be a task that modern AI is particularly well-suited to \cite{Wang2017,Piotrowski2019,Kaliszyk2018,Lederman2018,Kaliszyk2017,Alemi2016}. Improved decision-making heuristics in automated reasoning are especially important in proofs using non-classical formalisms, such as second-, higher-, or meta-logics. Such logics can sometimes allow for the expression of complex proofs in far fewer steps than might be required in a first-order logic \cite{Buss1994,Smith2007}. However, this increased expressive power also considerably expands the search space of any proof done in such logics, mandating the need for said improved heuristics. 

However, a platform to easily experiment with applying AI to a plurality of logical formalisms does not exist; at least not in a way that jointly satisfies desiderata that we will state shortly. In this paper, we will describe our progress in addressing these goals by presenting MATR (\textbf{M}achine \textbf{A}ccountability through \textbf{T}raceable \textbf{R}easoning), a new automated reasoning framework. As a proof-of-concept, we introduce a metalogic capable of expressing proofs of G\"{o}del's Incompleteness Theorems, and show how MATR can be used as a platform for developing AI systems capable of discovering and reasoning over such proofs. 

MATR is based on the following design principles:

    \Paragraph{P1. The underlying control system should be as formalism-agnostic as possible.} MATR began as an in-house tool to very quickly test the formal representations and inference rules related to variants of the Cognitive Event Calculus \cite{Arkoudas2009b,Bringsjord2013,Licato2014c,Bringsjord2014a,Bringsjord2015d}, whose visual style was inspired by the diagrammatic, flowchart-like aesthetic of Slate \cite{Bringsjord2008b} and the indented subproofs of Fitch-style natural deduction \cite{Barker2011}. Instead of creating an automated theorem prover from scratch for each new formalism, it was decided that a more flexible framework with easily interchangeable parts would be a better long-term strategy. 
    
    \Paragraph{P2. Semantics should be contained in the codelets and other interchangeable parts.} Any actions requiring semantic understanding of the contents of the nodes in MATR should be contained in one of MATR's interchangeable parts, preferably its codelets. \textit{Codelets} are independently operating programs which perform the bulk of the work in MATR, and are described more later in this section. Syntax checking, carrying out inference rules, recording type information, and even knowing whether a proof is completed are tasks delegated to individual codelets. This is also meant to enable rapid deployment and testing of nontraditional logics (e.g. higher-order, modal, inductive, informal, etc.). One trade-off of this flexibility is that it is entirely possible for a set of codelets to be mutually incompatible. Accordingly, MATR also allows for pre-built \textit{configurations} to be loaded in the form of a YAML text file. For example, if one wishes to use MATR as a natural deduction reasoner for standard first-order Peano Arithmetic, such a configuration will already be available to load.
    
    \Paragraph{P3. Codelets must be programmer-friendly, allowing for easy implementation and changes of inference rules without modifying the core.} Although much of MATR's code is written in Clojure, its codelets can be written in virtually any language (most are in Python).
    
    \Paragraph{P4. Control decisions and optimization strategies should be as interchangeable as possible.} Another motivator for MATR is as a platform for applying advances in Artificial Intelligence and Machine Learning to Automated Reasoning. For example, given partially completed proofs, a fruitful research question may be how to select which codelets to run, and over which existing formula nodes, in order to optimally complete the proofs. MATR encourages this by making two components interchangeable: the \textit{codelet chooser}, which selects which codelets to execute and the order to execute them each iteration;
    % \TODO{phrasing is confusing here}
    and the \textit{recommendation resolver}, which decides which codelet recommentations to actually apply to the proof.
    
    \Paragraph{P5. The front-end should be modular to satisfy a variety of use cases.} There are at least two primary use cases of MATR's user interface: for students relatively new to automated reasoning, and for researchers who may want to use MATR as part of a larger AI system. For the former case, we have an interactive graphical user interface; this is the source of the figures in this paper. For the latter, MATR's back-end can communicate with a lightweight command-line interface or interact directly with programs through an API.

\subsection{The Proof Space}

\begin{figure}[t]
\centering
\includegraphics[width=0.99\columnwidth]{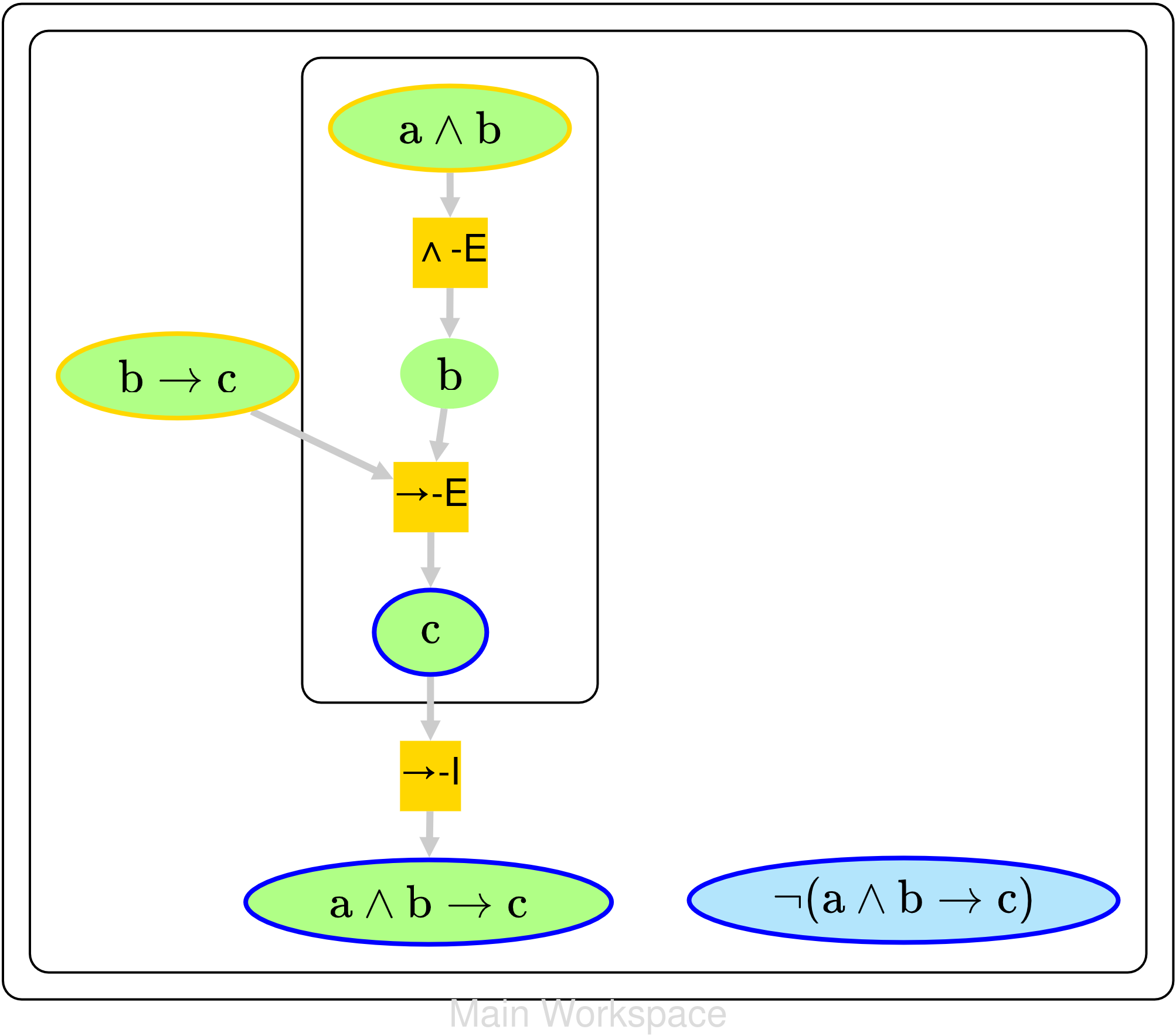}
\caption{Example simple proof that $(A \wedge B) \rightarrow C$ follows from $B \rightarrow C$. Formula nodes are ovals, while inferences are rectangles. Nodes start out as light blue and are colored green as they receive the checked flag. The axioms and goals of each box receive a border, colored gold or blue (resp.).}
\label{fig:simproof}
\end{figure}

    MATR's basic organizational unit is called the \textit{box}. A box may contain a set of formula nodes, inference nodes, other boxes, and connections between nodes or boxes. If a box $\mathcal{B}$ contains a node or box $o$, then we say that $o$ has $\mathcal{B}$ as a \textit{parent box}. All elements in MATR have exactly one parent box, with one exception: the top-level box. It is also referred to as the \textit{proof space} or \textit{root box} and is the starting point of all MATR proofs.
    
    Formula nodes contain formula objects, which are typically S-expression strings. Inference nodes are used to link a set of formula nodes $\mathbf{F}$ (the premise node) to another formula node $\phi$ (the conclusion node), where $\phi \notin \mathbf{F}$. Inference nodes contain information about the nature of the inference from $\mathbf{F}$ to $\phi$, typically a string corresponding to the name of the inference rule, metadata about which codelet suggested the inference, and other information of use to codelets. Thus, directed edges connect formula nodes to inference nodes, and inference nodes to formula nodes, but never directly connect two nodes of the same type.
    
    Formula nodes can also be \textit{axioms} or \textit{goals} of their parent box. If $\alpha$ is an axiom of box $\mathcal{B}$, then it is an axiom of any box which has $\mathcal{B}$ as a parent box. Formula nodes may also contain metadata such as the binary \textit{checked flag}, which denotes whether a formula node follows from the axioms of its parent box. Formula nodes which are axioms of their parent boxes automatically have checked flags, and a special codelet \textit{propagates} checked flags by looking at inference nodes: if, for inference node $i$, (1) all of $i$'s premise nodes have checked flags, and (2) $i$ is marked as a \textit{deductive}\footnote{Although not used in this paper, inference nodes can correspond to non-deductive inferences, in order to open the door to proof-solving heuristics and proof tactics.} inference, then $i$'s conclusion node is given a checked flag.
    
%    All of these can be seen in Figure \ref{fig:simproof}, where a proof is shown that $(A \wedge B) \rightarrow C$ follows from $B \rightarrow C$.  \TODO{the sample image should have some nodes checked and some unchecked, and we should explain here what checked nodes look like (e.g., color)}

\subsection{Core Components}
    The most important component of MATR's design is the \textit{codelet}, an individual piece of specialized code whose job is to analyze the current proof space, and make recommendations about how to modify the proof space (typically the addition of inference and formula nodes). In keeping with the design principles listed earlier, codelets can be written in virtually any programming language, are easily interchangeable, and are principally responsible for both defining and behaving in accordance with the intended semantics of the proof state. The term `codelet' is borrowed from the Copycat model of analogical reasoning \cite{Hofstadter1984,Mitchell1993,Hofstadter1995}, and like Copycat codelets, our codelets should be thought of as independently operating ``worker ants" which each specialize in searching for a unique set of features in the proof space (or in a local subset of the proof space), and making recommendations based on their findings.
    
    Most codelets directly correspond to inference rules; e.g., the \textit{Modus Ponens} codelet simply looks for formulae pairs of the form $\phi$, $\phi \rightarrow \psi$ and recommends adding a ``Modus Ponens" inference node and a $\psi$ formula node. Other codelets are responsible for background or support tasks, such as propagating checked flags, or ensuring that every formula node contains syntactically valid formulae. If the application calls for it, codelets may even contain entire external programs or automated theorem provers; e.g., we make use of a full first-order resolution codelet later in this paper.
    
    MATR's codelets typically operate either in a \textit{forward} or \textit{backward} mode. Forward-operating codelets look for sets of formulae that jointly satisfy some inference schema's premises, and suggest the addition of a conclusion, as in the \textit{Modus Ponens} example earlier. It is beneficial for such codelets to restrict its search to formulae that have checked flags, so that any new formula nodes added will also have checked flags propagated to them. Backwards operating codelets work in the other direction, adding formulae that lead to nodes which have yet to be checked. Allowing some codelets to operate unconstrained could result in a potentially infinite number of recommendations per iteration, such as a forward operating Disjunction Introduction codelet. This explosion is managed by heuristics in the codelet chooser and recommendation resolver.
    
    % using at least two strategies: First, allowing such codelets to operate in forward mode is treated as a sort of last resort by the Codelet Manager. Second, codelets such as Disjunction Introduction are, by default, used in backwards mode, as it seeks formulae nodes of the form $\phi \vee \psi$ which do \textit{not} contain checked flags, and recommends adding the formula nodes $\phi$ and $\psi$ above the existing formula node with corresponding inference nodes. Thus, if $\phi$ is given a checked flag later, that flag will propagate to $\phi \vee \psi$.
    
    The MATR \textit{core} (sometimes referred to as the \textit{back-end}) is the fixed central unit that coordinates all of MATR's interchangeable parts. In its simplest form, it merely serves as the conduit through which the components of MATR communicate and maintains the repository of shared proof state. Currently, core presents a REST API which wraps a datascript graph database that holds all of the proof state. Core also provides provides a minimal \textit{Codelet Chooser} and \textit{Recommendation Resolver}. The former can be configured to trigger codelets based on the result of queries against the core database, while the latter is mostly focused on sanity checking (e.g. preventing the addition of duplicate nodes and boxes).

    The \textit{front-end} is the primary way to interact with MATR as a whole. It wraps some of core's API into a graphical user interface that allows the user to upload a configuration to core, set up the axioms and goals of the proof space, trigger the execution of codelets, and view the resulting proof. 
    
    \textit{Codelet Servers} act as hosts for the codelets provided by core. They present REST endpoints for the codelet chooser to send messages to trigger the execution of particular codelets. The codelets can then make additional queries against core, perform logical inferences, and the codelet server responds to the request with actions for the recommendation resolver.
    
    % The \textit{Codelet Server} (implemented in python) acts as the host for various codelets, and the mechanism for which core communicates its queries to endpoints that, in turn, refer back to the codelets that are registered under those endpoints. The codelet server, when calling the individual codelets, does so in a multi-process fashion, allowing for each codelet to operate on their requisite data in isolate and reduce the overall computational time of applying the various codelets to a possibly large volume of formulas.

\section{Discovering the Proofs}

%\begin{figure*}[t]
%\centering
%\includegraphics[width=0.99\textwidth]{graphics/proof_whiteboard.jpg}
%\caption{Our sketch of the incompleteness proof, which we used as a guide for the proof in MATR}
%\label{fig:proof}
%\end{figure*}

\begin{figure*}[t]
\centering
\includegraphics[width=0.99\textheight, angle=90]{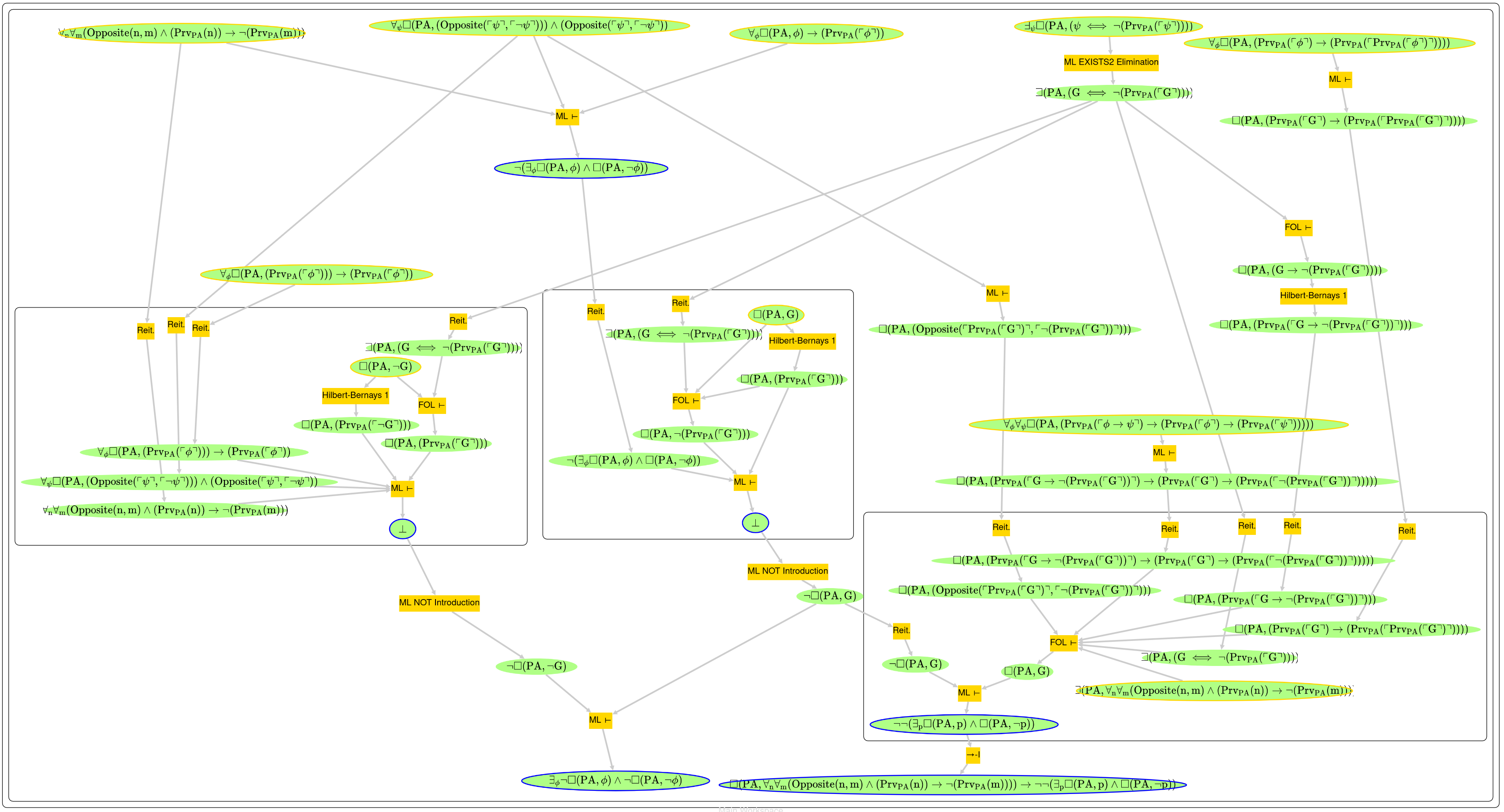}
\caption{MATR-discovered proof of both incompleteness theorems, after pruning and rearranging nodes to better make use of space}
\label{fig:fullproof}
\end{figure*}

\subsection{A Formalized Metalogic}
    As a demonstration and test of MATR's proof-finding capabilities, we set out to generate a proof of G\"odel's First and Second Incompleteness Theorems, given a proof sketch. Verifying proofs of these theorems are often used as a sort of stress test for an automated reasoner's formalism, and the the ability to automatically discover these proofs has been claimed by many with various degrees of success \cite{OConnor2005,Sieg2005,Licato2013a,Paulson2014,Paulson2015}. All existing machine-discovered proofs of the incompleteness theorems rely on carefully selected symbols and inference rules. An ongoing goal of such work is to discover proofs of the incompleteness theorems with increasingly minimal assumptions and custom tailoring of the prover's starting conditions. Here, our contribution does so with a formalism and automated reasoner robust enough to not only find concise proofs of both incompleteness theorems, but of similar theorems in the metalogical space as well.
    
    Our formalizations of the incompleteness proofs are loosely based on the versions and terminology described by \cite{Smith2007}. Let us assume a G\"{o}delian numbering scheme that assigns a unique integer to all formulae in first-order Peano Arithmetic (PA), using the normal conventions. Then for any well-formed-formula (wff) $\phi$ in PA, $\ulcorner \phi \urcorner$ denotes its corresponding G\"{o}del number. % in $\mathds{N}$.
    Slightly abusing notation for convenience, if $\Phi$ is a proof in PA, $\ulcorner \Phi \urcorner$ denotes its corresponding encoding (called a \textit{Super} G\"{o}del number). Given an integer in $n \in \mathds{N}$, determining whether $n$ encodes a PA wff, a well-formed proof in PA, or neither, is decidable and captured in PA itself \cite{Smith2007}.
    
    $\Prv(n)$ is shorthand for the wff formula in PA which expresses (but does not capture) the numerical property ``$n$ corresponds to the G\"{o}del number of a formula that is a theorem of PA.'' This predicate also satisfies the Hilbert-Bernays provability conditions (Equations \ref{eq:HB1} - \ref{eq:HB3} below).
    % , so that:
    % \begin{itemize}
    %     \item If PA $\vdash \phi$, then PA $\vdash \Prv(\QB{\phi})$
    %     \item PA $\vdash \mathit{\Prv(\QB{\phi \rightarrow \psi}) \rightarrow (\Prv(\QB{\phi}) \rightarrow \Prv(\QB{\psi}))}$
    %     \item PA $\vdash \Prv(\QB{\phi}) \rightarrow \Prv(\QB{\Prv(\QB{\phi})})$
    % \end{itemize}
    $\Opp(m,n)$ is shorthand for the PA wff which is provable if and only if $n$ encodes a PA wff which is the negated form of the PA wff that $m$ encodes. Because this formula trivially captures a primitive recursive property, it is captured in PA \cite{Smith2007} and thus it is an axiom that for all PA wffs $\phi$, PA $\vdash \Opp(\QB{\phi}, \QB{\neg \phi})$. Consistency in PA is expressed by the formula $\Con$:
    \begin{equation}
        \forall_{m,n} (\Opp(m,n) \wedge \Prv(m)) \rightarrow \neg \Prv(n)
    \end{equation}
    
    %Note that both $\Opp$ and $\Con$ express primitive recursive functions, and are thus both captured / decidable in PA \cite{Smith2007}. %why did I write this? \Con isn't primitive recursive.
    
    \noindent Often, explanations of the incompleteness proofs will use high-level natural language which appears to quantify over formulae---e.g., ``there exists a formula $\phi$ such that $\phi$ has a certain property." This, we believe, reflects the natural relative ease of reasoning about formula as discrete objects which can have properties, at least when working through the proofs of the incompleteness theorems. Our formalization of the incompleteness proofs thus uses a first-order modal typed metalogic, to more directly reflect the kinds of statements used in explanations of the proofs. The objects of our metalogic are the same as in $PA$, with the addition of two types: Proof theory and formula symbols. Proof theory symbols enable metalogical formulae that compare and express the properties of entire proof theories, but for this paper we only use one: $PA$. Formula symbols are objects corresponding to PA wff. Since the metalogical version of the universal and existential operators can quantify over formula symbols, the metalogic allows us to write formulae such as Equations \ref{eq:HB1} through \ref{eq:G2} below.
    
    More precisely, a metalogic wff is either: (1) a single formula symbol, (2) a PA wff, or (3) if $\phi$ and $\psi$ are metalogic wffs, then metalogic wffs include: $\neg \phi$, $\phi \vee \psi$, and so on using the normal rules of logical operators. All operators in first-order PA have analogs in the metalogic. Only one additional operator is added: The provability operator $\vdash$ is written $\square(p,f)$, where $p$ is a proof theory, and $f$ is a metalogic wff. It should be read, ``the proof theory $p$ has the formula corresponding to $f$ as one of its theorems.'' Unlike $\Prv(\QB{\psi})$, which is a PA wff, $\square(PA, \psi)$ is a metalogic wff. Furthermore, $\forall_\phi \square(PA, \phi) \rightarrow \Prv(\QB{\phi})$ and $\forall_\phi \square(PA, \Prv(\QB{\phi})) \rightarrow \Prv(\QB{\phi})$ are provided as axioms, but the converse are not.
    
    Given this notation, we can introduce the three Hilbert-Bernays provability conditions as axioms as well:
    \begin{equation} \label{eq:HB1}
        \forall_\phi \square(PA, \phi) \rightarrow \square(PA, \Prv(\QB{\phi}))
    \end{equation}
    \begin{equation} \label{eq:HB2}
      \begin{aligned}
        \forall_{\phi,\psi} & \square(PA, \mathit{\Prv(\QB{\phi \rightarrow \psi})} \\
        & \mathit{\rightarrow (\Prv(\QB{\phi}) \rightarrow \Prv(\QB{\psi}))})
      \end{aligned}
    \end{equation}
    \begin{equation} \label{eq:HB3}
        \forall_\phi \square(PA, \Prv(\QB{\phi}) \rightarrow \Prv(\QB{\Prv(\QB{\phi})}))
    \end{equation}

    \noindent Primarily as a matter of notational convenience, we define an additional definition of consistency $Con(PA)$:% 
    
    \begin{equation}
        \neg \exists_\phi \square(PA,\phi) \wedge \square(PA, \neg \phi)
    \end{equation}
    
    \noindent Note that $Con(PA)$ is a metalogic wff, whereas $Con_{PA}$ is a PA wff. $Con(PA)$ also draws on a notation which emphasizes $PA$ as an object that can be replaced with other formal theories.\footnote{By this sort of design we intend, in the future, for MATR to reason about metalogical properties of multiple formal theories within the same proof.} In any case, $Con(PA)$ and $Con_{PA}$ can be derived from each other in our present proof space, and only the latter is used as an axiom. Thus, we can concisely formalize G\"{o}del's two Incompleteness Theorems as follows:
    \begin{equation} \label{eq:G1}
        \exists_\phi \neg \square(PA, \phi) \wedge \neg \square(PA, \neg \phi)
    \end{equation}
    \begin{equation} \label{eq:G2}
        \square(PA, \Con) \rightarrow \neg Con(PA)
    \end{equation}

\subsection{Setup}
    
    To establish a basis for the proof, two initial setup codelets populate the proof space with several intermediate nodes and boxes to act as the foundation for later codelets. By statically introducing these intermediate nodes and boxes we guide the proof search process in a way which avoids combinatorial explosion. While the front end could be used to introduce each of these boxes and nodes by hand, using setup codelets simplified the process of re-running the entire proof while testing. In the future we hope to use heuristics to completely automate this as well. In our proof sketch we used three subproofs, which are created in this step. They are handled in setup due to the need for intermediate nodes to be placed in those subproofs as well. Additionally, the existential elimination acting on one of the axioms occurs in the setup so that the symbol chosen to take place of the quantifier is fixed to be the same as those used in the intermediate nodes.
    % This is necessary because none of those subproofs can be resolved with a single resolution step- all three require a combination of resolution over first-order logic and resolution over the metalogic to derive a contradiction.
    
    % Talk about the setup codelet, creating the proof state, the intermediate nodes we introduced, justify the four "hardcoded" justifications
    
    In order to simplify the proof space, we employ four types of resolution codelets; two for the metalogic, and two for first-order logic. The two ``$\mathit{PA\vdash}$" codelets simply search for formulae nodes $\square(PA, \phi_1), ..., \square(PA, \phi_n)$ with checked flags, and either a formula $\square(PA, \psi)$ without a checked flag or a $\square(PA, \bot)$ formula. %Having both simplifies how we interface with our resolution decision procedure. 
    It then attempts to show that, subject to timeout and memory limitations, $\psi$ follows from $\{\phi_1, ..., \phi_n\}$ using first-order resolution without equality. Likewise, the ``$\mathit{ML\vdash}$" codelets search for formula nodes $\gamma_1, ..., \gamma_m$ with checked flags and uses resolution to prove whether a formula $\psi$ or $\bot$ without a checked flag follows. Although these codelets are quite powerful (and potentially combinatorially explosive), they do not operate in a purely forward mode, which severely limits what can be practically derived. But neither can either of these codelets complete the entire proof on their own. Consider the three Hilbert-Bernays Provability Conditions: After $\forall$-Elimination, Equations \ref{eq:HB2} and \ref{eq:HB3} produce subformulae that can be reasoned over by $\mathit{PA\vdash}$, but Equation \ref{eq:HB1} does not---it requires inferences at the metalogical level. Conceivably, one might simply strengthen $\mathit{ML\vdash}$ to subsume what $\mathit{PA\vdash}$ does, but this explodes its average run time, removes MATR's ability to introduce smarter heuristics, and would prevent us from using a general purpose first order resolution implementation.
        
    In order to ensure proper connection between the nodes introduced by the resolution codelet and the nodes that already exist outside of the subproof, a reiteration codelet connects the node outside of the subproof to the corresponding node inside the subproof. This allows for propagation of checks into the subproof, so that resolution can truly demonstrate that it has proven a particular formula with accordance to information both inside and outside of the subproof.
    
    We also implemented a dedicated codelet for Hilbert-Bernays 1 \eqref{eq:HB1} instead of providing it as an axiom of the root box. In testing, we discovered that providing HB1 as an axiom lead to an explosion of the number of necessary clauses to process in the resolution procedure. More specifically, the proof of Con(PA) with resolution explodes from 97 clauses to not resolving with an upper bound of 2000 clauses. The HB1 codelet works purely in a backwards mode, adding inferences for nodes of the form $\mathit{\square(PA, \Prv(\QB{\phi}))}$ and connecting to (or creating) a corresponding node of the form $\mathit{\square(PA, \phi)}$.

    % Talk about reiteration? Does it make sense to move this earlier in the paper?
    % Talk about check prop? We already introduced it at a high level

    % Therefore, we our ``$\mathit{PA OneStep}$'' codelet searches for formulae nodes $\square(PA, \phi_1), ..., \square(PA, \phi_n)$ with checked flags, converts all $\phi_i$ into clauses, and performs exactly one iteration of first-order resolution between each pair of clauses. All resolvents are then converted back into first-order formulae. In order to minimize the clutter, we discard any that are of the form $\alpha \vee \neg \alpha$. All remaining formulae $\psi$ are passed to the recommendation resolver as $\square(PA, \psi)$.
    
    Once the proof is completed, a pruning process identifies (and displays) the shortest proofs of each goal node in the root box. Figure \ref{fig:fullproof} shows this pruned proof. Pruning is accomplished in two steps. The first propagates forward from the axioms of every box to count the number of justifications in the shortest proof of any reachable and checked node. Given these proof size counts, a backwards traversal is performed from each goal of the root box to extract the nodes and boxes used in the shortest proof of that goal. 
    
    With the provided intermediate nodes, the proof takes four iterations to complete. Interestingly, the proof found by MATR is actually shorter than the proof sketch we used to select our intermediate nodes and high level proof structure. Most surprising to us was MATR using part of the proof of the first incompleteness theorem in its proof of the second, short circuiting three intermediate nodes.
    % Talk about the differences between the proof we generated and the one we had worked out by hand- using the (NOT (PROVES PA G)) of first incompleteness in second, skipping a few intermediate nodes, etc.

% \subsection{Robustness Testing and Analysis}
%     \TODO{explain what we did to test how robust it was in finding the proofs (1/2 page)}
%     - e.g., a table or graph how the number of total nodes created (before trimming) / run time changes as the number of inference rules are increased to include full FOL
%     \TODO{quantitative and qualitative analyses of the results - various observations}

\section{Conclusion and Future Work}
    With MATR, we sought to satisfy the design criteria \textbf{P1} - \textbf{P5} by designing a system that operates fluidly over various logical systems with a series of interchangeable components. MATR allows the use of various heuristics to verify or discover proofs with otherwise-explosive proof spaces. To demonstrate the potential of this system, we verified and simplified proofs of G\"{o}del's Incompleteness Theorems.
    
    We see a few directions to take MATR going forward. We have already made progress automating the generation of intermediate nodes in the incompleteness theorem proofs, but we are still exploring the heuristics needed to minimize explosion of the proof space. Remarkably, MATR discovered a shorter proof than we told it to find, which we consider a strong indicator of the potential for work in this area. As MATR is designed to reason in higher order logics, it may be immensely valuable in automatically discovering shorter proofs of already known theorems. %To this end we push forward.
    %Additionally, changes can be made to core to rectify potential failures as a consequences of a considerably large volume of noise propagating into the system. 
    %We also plan to apply MATR to a proof of G\"{o}del's Speedup Theorem, where we expect the graph representation of proofs to be especially valuable. 
    The full source code for MATR and what is necessary to replicate this proof will be available upon publication.

\subsection{Acknowledgements}

MATR owes its existence to work started at the Rensselaer AI and Reasoning (RAIR) Lab and contributed to by many students and collaborators since then---too many to reasonably name here. We are extremely grateful to the support and contributions of them all.

\textit{This material is based upon work supported by the Air Force Office of Scientific Research under award numbers FA9550-17-1-0191 and FA9550-18-1-0052. Any opinions, findings, and conclusions or recommendations expressed in this material are those of the authors and do not necessarily reflect the views of the United States Air Force.}

% \begin{figure}[t]
% \centering
% \includegraphics[width=0.9\columnwidth]{figure1} % Reduce the figure size so that it is slightly narrower than the column. Don't use precise values for figure width.This setup will avoid overfull boxes. 
% \caption{Using the trim and clip commands produces fragile layers that can result in disasters (like this one from an actual paper) when the color space is corrected or the PDF combined with others for the final proceedings. Crop your figures properly in a graphics program -- not in LaTeX}
% \label{fig1}
% \end{figure}

\bibliographystyle{aaai}
\bibliography{john}

\end{document}